\documentstyle[prl,aps]{revtex} %\def\btt#1{{\tt$\backslash$#1}} 
\draft
\begin{document}
\twocolumn[\hsize\textwidth\columnwidth\hsize\csname
@twocolumnfalse\endcsname

\title{Backscattering: an overlooked effect of General Relativity?} 

\author{Edward Malec}
\address{Jagiellonian University, Institute of Physics, {3}0-59 Krak\'ow, Reymonta 4, Poland} 
\author{Niall \'O Murchadha}
\address{Physics Department, University College, Cork, Ireland} 
\date{\today}
\maketitle 
\begin{abstract}
    The total flux of outgoing radiation  in a strong gravitational field  
   decreases due to backscattering if the sources are close to
 an apparent horizon. It can   cause detectable changes in the shape of signals.  
Backscattering could well be of relevance to astrophysics and  would constitute  a new test of the
validity of general relativity. An explicit bound for this effect is derived for scalar fields.

\end{abstract}


\vskip2pc]
Backscattering  can be described as a phenomenon which causes the radiation of a massless field 
to disperse outside those null cones  defined by the initial impulse.
It is   known in older mathematical literature as the `breakdown of Huyghens principle'
\cite{Hadamard} and it is interesting for a number of reasons. 
First, as a non-Minkowski spacetime  effect, it offers  a new way of testing
general relativity. Second, it can be important in astrophysics, both in order to
explain the radiation coming to us from regions adjoining  a black hole and to infer
information about the sources of the radiation. Third, a strict upper bound on the magnitude of 
the backscattering effect, such as we derive here, offers  numerical relativists an independent 
test of the correctness of their numerical codes.
  
   Our model is a  spherically symmetric massless
scalar field emitted from a region near to, but outside, a non-rotating black hole.  These
simplifying assumptions should not seriously restrict the validity of our conclusions.   
For example, the propagation of electromagnetic fields, we believe, should obey inequalities 
similar to the ones derived below. These estimates, while they   break down close to the
horizon, allow us to distinguish the region in which the backscattering may play a significant
role from that in which it is of no importance.  They offer, to our knowledge, the first
quantitative measure of this strong field effect. Others who work on backscattering adopt quite
different approaches \cite{NS}. 
  
 We consider a foliation of the spacetime by using the polar gauge
slicing  condition, $tr K = K_r^r$; that is, with a diagonal line element
\begin{equation} ds^2 = - \beta (R, t)\gamma (R, t)dt^2 + {\beta (R, t)\over \gamma (R, t)} dR^2 +
R^2(r, t) d\Omega^2~,
 \label{4.1}
 \end{equation} 
where $t$ is the time, $R$ is a radial
coordinate which coincides with the areal radius and $d\Omega^2 = d\theta^2 + \sin^2\theta d\phi^2$
is the line element on the unit sphere, $0\le \phi < 2\pi $ and $0\le \theta \le \pi $;   
  $\beta$ and $\gamma$ go to $+1$ at infinity. 

  For a  massless scalar field   the stress-energy  is
$T_{\mu \nu }= \nabla_{\mu }\phi \nabla_{\nu }\phi -g_{\mu \nu }\nabla_{\alpha }\phi
\nabla^{\alpha }\phi/2 $. The matter energy density is $\rho =-T_0^0 $ and the matter current
density is $J = -T_{0R}/\beta$.  $(\partial_0 \pm \gamma\partial_R )$ are the outgoing and 
ingoing null directions. We define  radiation amplitudes  
\begin{eqnarray}
h_+(R,t) &= h(R, t) &={1\over 2\gamma} (-\partial_0 +\gamma\partial_R ) (R\phi )\\
h_-(R,t) &= h(-R, t)&={1\over 2\gamma} (\partial_0 +\gamma\partial_R ) (R\phi ).
\label{4.2}
\end{eqnarray}
One can show \cite{Malec1997} that
\begin{equation}
\beta (R) = e^{-8\pi \int_R^{\infty } {dr\over r} \left[( h_+ -\hat h)^2 + 
(h_- -\hat h)^2\right]}~,
\label {4.3}
\end{equation}
and
\begin{equation}
\gamma(R) = 1 - {2m_0 \over R} + {1 \over R} \int_R^{\infty}[1 - \beta (r)]dr~, 
\label{5.2}
\end{equation}
where $\hat h= -{1\over 2R}\int_{R}^{\infty}dr[h_+(r) + h_-(r)]={1\over 2}\phi~.$ 
$\gamma $ can be expressed in the following useful form \cite{Malec1997}
\begin{equation}
\gamma (R) = \left( 1- {2m_0\over R} +{ 2 m_{ext}(R)\over R}\right) \beta (R) 
\label {4.8}
\end{equation}
where $m_0$ is the asymptotic (total) mass and $m_{ext}$ is a contribution to 
the asymptotic mass coming from the exterior of a sphere of a radius $R$, 
\begin{equation}
m_{ext}(R) = 4\pi\int_{R}^{\infty }{\gamma \over \beta } \left( [h_+(r)-\hat h ]^2+
[h_-(r)-\hat h ]^2\right) dr~. \label {4.9} 
\end{equation}
The scalar field equation
$\nabla_{\mu }\partial^{\mu } \phi =0 $
can be written as a single first
order equation on a `symmetrized' domain $-\infty \le R\le \infty $ \cite{Malec1997} by writing
$h(R) = h_+(R)$    and $h(-R) = h_-(R)$ as
\begin{equation}
(\partial_0 +\gamma \partial_R)h = (h -\hat h ) {\gamma -\beta \over R}~.
\label {4.6}
\end{equation}
 Eq. (\ref{4.6}), together with the definitions of $h, \hat h , \beta $, and $ \gamma $,
 is equivalent to the Einstein equations coupled to the scalar field. 
 The external mass changes along an outgoing null cone according to
\begin{equation}
(\partial_0 + \gamma\partial_R)m_{ext}(R) = -8\pi\gamma^2 (h_- - \hat h)^2 ~.
\label{4.10}
\end{equation}
The polar gauge allows us
 to express the metric directly in terms of the matter, Eqs. (\ref{4.3} - \ref{5.2}), where
  all the integrals are in the exterior region.  
The local and global Cauchy  problems for the above system  are solvable in an external region
bounded from the interior by a null cone (\cite{Malec1996b},
\cite{Malec1997}). 

 Following Eq.(\ref{4.8}),  $(\gamma -\beta)/R= -2m(R)R\beta/|R|^3$ where
   $m(R) = m_0 - m_{ext}(R)$ is the Hawking mass at a radius $R$.  
  Eq.(\ref{4.6}) gives a `red-shift' due to the $h$ term on the right-hand-side
(determined by the mass function, $m(R)$, rather than the Schwarzschild mass,
$m_0$) and a `backscattering' due to the $\hat h$ term. 

Let us define
\begin{equation}
\ln\left[1 - {2\tilde{m}(R) \over R}\right] = -\int_R^{\infty}{2m(r) dr 
\over r^2(1 - 2m(r)/r)}~,
\label{4.13}
 \end{equation}
where the integral is taken along an outgoing null ray. 
 This allows us to rewrite Eq.(\ref{4.6}) as
\begin{equation}
(\partial_0 + \gamma \partial_R)(1 - {2\tilde{m}(R) \over R})h =
\hat h (1 -  {2\tilde{m}(R) \over R}){2m(R)R\beta \over |R|^3}~. 
\label{4.15} 
\end{equation}
Thus   $(1 - 2\tilde{m}(R)/R)$ is the redshift factor and the
right-hand-side of Eq.(\ref{4.15}) determines the backscattering.
 
It is natural to write Eq.(\ref{4.3}) as
\begin{equation}
\beta (R) = e^{-8\pi \left(\int_R^{\infty }+\int_{-R}^{-\infty }\right) 
{\beta \over \gamma r} {\gamma \over \beta}( h -\hat h)^2 dr }~, 
\label {5.4} 
\end{equation}
and by using Eq.(\ref{4.8}) we get that
\begin{equation}
{\beta \over \gamma r} = {1 \over r - 2m(r)} \le {1 \over r - 2m_0}~.
\label{5.5}
\end{equation}
The factor $\beta/\gamma r$ can be taken  out of the integral in Eq.(\ref{5.4}) 
and replaced with its value at $r = R$. The remainder is then essentially right hand side of
Eq.(\ref{4.9}). Let us choose an $\epsilon$  and an $R_A$ such that
$m_{ext}(R_A)/m_0 < \epsilon$ and $R_A > 2m_0(1+\epsilon)$. For any $R \ge R_A$ we obtain
\begin{equation}
1 \ge \beta(R) \ge \beta(R_A) \ge e^{-2{m_{ext}(R_A) \over R_A - 2m_0}}  
  \simeq 1 - O(m_{ext}/\epsilon m_0)~.
\label{5.6}
\end{equation}
In the same vein, using Eq.(\ref{4.8}) one gets $\gamma (R) \simeq 1 - 2m_0/R +
O(m_{ext}/\epsilon m_0)$. Thus the effect of the matter on the geometry can be controlled. Notice
that    in regions sufficiently close to  the horizon, even if $m_{ext}/m_0 \ll 1$, a small
cloud of matter can still strongly influence the geometry. In what follows, however, we will
restrict our attention to the region outside  $R=3m_0$, since only in that region of spacetime can
we get sensible analytic estimates. This is equivalent to choosing $\epsilon \ge 1/2$.

 For $R>3m_0$  the scalar wave equation Eq.(\ref{4.15}) can be approximated as 
  \begin{equation}
(\partial_0 +\gamma \partial_R)(1 - {2m_0 \over R})h = \hat h ( 1 - {2m_0 \over R})
{2m_0R \over |R|^3}~,
\label {5.8}
\end{equation}
 with the error terms   of order $m_{ext}/m_0$. 
 In the limit of $m_{ext}/m_0 \ll 1$   the   equation  describes
  a scalar field propagating on a fixed Schwarzschild background.

One can show, improving a coefficient in an inequality of \cite{Malec1997}, that 
\begin{equation}
|\hat h|\le {\sqrt{ m_{ext}}\over R^{1/2} \sqrt{8\pi (1-{2m_0\over R})}}.
 \label {5.9}
\end{equation}
(A similar estimate  with $m_0$ rather than $m_{ext}$, appears in 
\cite{Demetrios}.) 
  Let us introduce the Regge-Wheeler coordinate \cite{MTW} 
$ r^* = R+2m_0 \ln (R/2m_0 - 1) $
so that $\gamma \partial_R = \partial_{r^*}$. The solution of Eq.(\ref {5.8}) can be
estimated above and below by solutions of 
\begin{equation}
(\partial_0 + \partial_{r^*}) \Bigl( (1-{2m_0\over R})h \Bigr) =^+_- { 2m_0 (1-{2m_0\over R})
\sqrt{ m_{ext}}\over R^{5/2} \sqrt{8\pi (1-{2m_0\over R})}}.
 \label {5.10} 
\end{equation}
 Eqns.(\ref {5.10})    are solved by 
 \begin{eqnarray}
&\left(1-{2m_0\over R(r)}\right)h(r^* , t) = h_0(r^* - t)\nonumber\\& \pm \sqrt{{m_{ext}\over 2\pi
}}m_0
\int_{(r - t, 0)}^{(r, t)}{\left(1 - {2m_0 \over R}\right) \over R^2 \sqrt{ R-2m_0}}dv~,
\label {5.12}
\end{eqnarray}
here the $dv$ in Eq.(\ref{5.12}) is   $dr^*$,   with
$dR/dr^* = 1 -2m_0/R$. If we ignore the second term   then the standard redshift is obtained,
\begin{equation} h(r^*(0) + \tau, \tau) ={\left(1 - {2m_0 \over R(r^*(0))}\right)\over
 \left(1 - {2m_0 \over R(r^*(0) + \tau)}\right)} h(r^*(0), 0)~,
 \label{3.10} 
\end{equation}
  while the integral in Eq.(\ref{5.12}) can be solved to give
\begin{eqnarray}
&\sqrt{{m_{ext} \over 2\pi}}m_0
\int {dR\over R^2 \sqrt{( R-2m_0}} =\nonumber\\ & \sqrt{{m_{ext} \over 16 \pi m_0}}
\left[\sqrt{{2 m_0 \over R}\left(1 - {2m_0 \over R}\right)} + 
\arctan \sqrt{{R - 2 m_0 \over 2 m_0}}\right] ~. 
\label{5.15}
\end{eqnarray}

 Thus the total backscattering is bounded by a term of order
$\sqrt{m_{ext}/m_0}$. In the limit where $m_{ext}/m_0$ is small, we recover the
 usual gravitational redshift  along an outgoing null ray  with an error of order 
$\sqrt{m_{ext}/m_0}$. This holds even in a strong gravitational field and  no
`quasi-static' assumption need be made. We need not assume $R \gg m_0$, but only $R \ge 3 m_0$. 
 
Let us assume that   initial data    at $t = 0$  represents a pure outgoing wave.
In other words,   $h_-(t = 0) \equiv 0$.
We select an $R_A \ge 3m_0$ on the initial slice such that $m_{ext}(R_A)/m_0 \ll 1$. Finally we
assume, in addition  to $\phi \rightarrow 0$ at infinity, that $\phi = 0$  for $R < R_A$. This
guarantees that the radiation is bounded away from the black hole. Consider the future outgoing
lightray from
$(R_A, 0)$. This is well outside any event  horizon, which would be at approximately $R = 2m_0$, so
this lightray is really outgoing all  the way to null infinity. Note that $m_{ext}(R_A, 0)$ is the
maximum value of
$m_{ext}$ over the whole wedge bounded by the initial slice and the outgoing null ray.
 We will estimate $h_-$ and $\hat h$
along this null ray. The  integration of 
Eq.(\ref{4.10}) along this ray  yields an estimate of the total
energy flux across this surface in the inward direction. This will be the total energy 
loss from the outgoing wave due to backscatter.

Choose a point on the null ray from $R_A$ and
label it by $(R_1, T_1)$.  To calculate $h_-$ at this point, consider
the ingoing future null ray which passes through this point and integrate Eq.(\ref{5.10}) along
this ingoing ray. This will start from  the initial hypersurface at some point $(R_2, 0)$ with $R_2
> R_A$. Along this null ray $R$ monotonically decreases while
$m_{ext}$ monotonically increases.    

To get an explicit estimate,  the integral in Eq.(\ref{5.12}) can be further approximated;
since $R$ monotonically decreases along the ingoing lightray, we can
replace the $\sqrt{1 - 2m_0/R}$ by $\sqrt{1 - 2m_0/R_1}$.  This yields 

\begin{eqnarray}
& 
\int_{R_2}^{R_1} {dR\over R^2 \sqrt{ R-2m_0}} \le  \nonumber \\ &   \sqrt{{R_1 
\over R_1 - 2m_0}}  \int {dR \over R^{5/2}}  
 \le{2 \over 3 R_1^{3/2}}\sqrt{{R_1 \over (R_1 - 2m_0)}}  ~.
\label{6.4} 
\end{eqnarray}
Thus we arrive at
\begin{equation}
|\gamma h_-(R)| \le {2\over 3} \sqrt{{m_{ext} \over 2\pi m_0}} 
\left[{m_0^{3/2} \over R\sqrt{R - 2m_0}}\right]~. 
\label{6.6}
\end{equation}

We can write, using  Eq.(\ref{4.2})  and  (\ref{6.6}),
\begin{equation}
(\partial_0 + \partial_{r^*})(R\hat{h}) = \gamma h_- \le {2 \over 3} 
\sqrt{{m_{ext} \over 2\pi m_0}}\left[{m_0^{3/2} \over R(R - 2m_0)^{1/2}}\right]
\label{6.7}
\end{equation}
and a similar inequality with a minus sign to give a lower bound of $\hat h$. 

These equations are integrated along the outgoing null ray from $R_A$ to give  
\begin{equation}
|R\hat{h}(R,t)| \le |R\hat{h}(R_A,0)| +
{2 \over 3} \sqrt{{m_{ext} \over 2\pi }}m_0 \int{dR \over (R - 2m_0)^{3/2}}~,
\label{6.8}
\end{equation}
Since we demand that $\phi(R_A, 0)  = 0$,
the first term in Eq.(\ref{6.8}) vanishes.
Thus we can bound $\hat{h}$ by
\begin{eqnarray}
&|\hat{h}(R,t)| \le  
{4 \over 3}\sqrt{{m_{ext} \over 2\pi m_0}}{m_0^{3/2} \over R(R_A - 2m_0)^{1/2}} -\nonumber \\ &
{4 \over 3} \sqrt{{m_{ext} \over 2\pi m_0}}{m_0^{3/2} \over R(R - 2m_0)^{1/2}}~. 
\label{6.8a}
\end{eqnarray}
 The last term in Eq.(\ref{6.8a})
is strictly larger than $|h_-|$ as given by Eq.(\ref{6.6}) if $R > R_A > 4m_0$ (i. e., when {\it
all} radiation is placed outside $4m_0$).  Thus  
\begin{equation}
|\hat{h} - h_-| \le |\hat{h}| + |h_-| \le {4\over 3} \sqrt{{m_{ext} 
\over 2 \pi m_0}}{m_0^{3/2}R_A \over R(R_A - 2m_0)^{3/2}} ~. 
\label{6.9}
\end{equation}

Therefore, from Eq.(\ref{4.10})   the total change in $m_{ext}$  satisfies 
\begin{equation}
\Delta m_{ext} \le
m_{ext} {16\over 9}
\left( {2m_0 \over R_A} \right)^2
\left({1 - m_0/R_A \over 1 - 2m_0/R_A}\right)~.\label{6.10} \end{equation}

From this expression we can see how sensitive the amount of backscattering 
is to the location of the innermost null cone. This estimate becomes meaningless if $R_2 \approx
3.5m_0$ because  $\Delta m_{ext}>m_{ext}$. However if $R_A = 6m_0$ we get that less than 25\% of
the total energy in the exterior field is backscattered. In the case of a neutron star, where
$2m_0/R \le 0.1$ (on the surface of the star), we have an upper bound for the backscattered energy
of 2\% of $m_{ext}$. 

In deriving the bound Eq.(\ref{6.10}), a number of truncations and approximations were used. While
most of them are sharp in the sense that configurations exist that turn the inequality into an
equality, we do not believe that all of them can be sharp simultaneously. Therefore Eq.(\ref{6.10})
is clearly an overestimate. We present it here, not because it is the best that can be done with
this technique, but because it is simple to derive, it is in analytic form, and is easy to interpret
physically. 

 For example, in Eq.(\ref{6.4}) we ignored the term  depending on  $R_2$. We can include this term
if the  initially outgoing pulses are far enough from the apparent horizon
\cite{EMNOMTCH}, so as to sharpen the estimate to
$m_{ext}$  

\begin{equation}
\Delta m_{ext} \le
{ 16\alpha^2 \over 9} \Bigl( {2m_0\over R_2}\Bigr)^2 {(1 - m_0/R_2) \over 1-2m_0/ R_2}m_{ext}~, 
\label{6.20} 
\end{equation}
where $\alpha^2 \simeq 0.4$. 
Applying this to the case of neutron stars, when the conditions assumed above hold true, we find
that the maximal amount of backscattered radiation cannot exceed 1 percent.

Eq.(\ref{5.12}) can be integrated in closed form, (Eq.(\ref{5.15})), but in Eq.(\ref{6.4}) we
approximated it to get a simpler expression because we needed to integrate the result
again. Numerical integration would obviate the need for this approximation entirely, but the result
would be much less transparent.

 The tail term can also be bounded using this method. The initially outgoing field, $h_+$,
generates a weaker ingoing field, $h_-$, which enters the `no-radiation' zone behind the
wavefront. This, in turn, scatters again off the gravitational field to generate a new
outgoing field, which turns up at null infinity at a later time, after the first burst
of outgoing radiation has
gone. This is the so-called  `tail' term. Eq.(\ref{6.8a}) gives us an estimate 
for $\hat{h}$ in the `no-radiation' zone. Substituting this back into the scalar wave equation
(\ref{5.8}), an estimate of the second order $h_+$ along the outgoing null ray is
\begin{eqnarray}
&|\gamma h_+(R,t)| \le    \nonumber \\ &
{4 \over 3}\sqrt{{m_{ext} \over 2\pi m_0}}{m_0^{5/2} \over \sqrt{R_A - 2m_0}} 
\int_{R_A}^R \left[ {1 \over R^3} - {\sqrt{R_A - 2m_0}\over R^3\sqrt{R - 2m_0}}\right]dR~.
\label{6.12}
\end{eqnarray}
In the limit as $R \rightarrow \infty$ along the null cone we get
 \begin{equation}
|h_+(\infty,\infty)| \le {4 \over 9}
\sqrt{{m_{ext} \over 2\pi m_0}}{(m_0/R_2)^{5/2} \over \sqrt{1 - 2m_0/R_2}}~. \label{6.14}
\end{equation}
This should be compared with the leading term in (\ref{5.12}), $h_0$. Using the
definition of the external mass $m_{ext}$, we get
 that the tail term is smaller than the leading term by a factor $(m_0/R_0)^2$.

 We expect that photons  will behave in a way similar 
to the massless spin zero field that we have  discussed here. Gravitational physics
predicts two phenomena for a radiating plasma surrounding a black hole or a neutron star:
Redshift  diminishes the intensity and frequency of the outgoing radiation while the total
energy in the radiation (as measured by the mass function) remains unchanged. Backscattering,
the other effect,  weakens the overall outgoing  radiation,   
so that the total energy that reaches infinity is reduced. In addition, backscattering changes
the shape of   extended signals - the leading part weakens while the rest of the impulse
gains in intensity. This conclusion follows from a careful analysis of  (\ref{4.6})
and is supported by   the  numerical results of \cite{EMNOMTCH}, which 
 show that up to $10\%$ of the radiation emitted at
$R = 3m_0$ is   `shifted' inside the main impulse; even if it reaches infinity, it does so with a
significant delay.

There are two astrophysical situations where those two different aspects of backscattering
will play a role. The net 
efficiency of the black hole - matter system   is reduced below what is
expected when only the redshift factors are taken into account. This  may   be of significance 
in modelling `too faint' galactic  nuclei fuelled  by black holes, such as the nuclei of M87 
\cite{Rey96}.   Second, backscattering may    leave imprints in X ray bursts \cite{Lewin} resulting
from energetic processes on the surface of a neutron star or in  collisions of compact bodies, such
as neutron star - neutron star,   neutron star - black hole, or black hole - planetoid mergers.
  Backscattering will deform the peak and contribute a tail term  to the radiation emitted from such
short-lived sources.   Gamma ray bursts (GRB) are believed by some \cite{p95} to arise from 
such collisions and should reveal   traces of  backscattering. The absence of these
effects would give great support to the   fireball  scenario of GRB's \cite{p97}.

Another more immediate application is in numerical relativity.  Much work has been done in
constructing codes to analyse the Einstein - massless scalar field model \cite{choptuik}. Bounds
on the magnitude of  the backscattering effect, such as derived here,  would
offer  numerical relativists a reliable test of the correctness  and long-time stability of their
codes. 
 
\acknowledgements{This work has been partially supported by the Forbairt grant SC/96/750 and
the KBN grant 2 PO{3}B 090 08.}

\end{document}